\documentclass[aps,prb,groupedaddress,notitlepage]{revtex4-1}

\usepackage{graphicx}% Include figure files
\usepackage{amsmath}
\usepackage{braket}
\usepackage{appendix}
\usepackage{bbm}

\begin{document}

% Use the \preprint command to place your local institutional report
% number in the upper righthand corner of the title page in preprint mode.
% Multiple \preprint commands are allowed.
% Use the 'preprintnumbers' class option to override journal defaults
% to display numbers if necessary
%\preprint{}

%Title of paper
\title{Approximating the frequency dependence of the effective interaction in the functional renormalization group for many-fermion systems}

% repeat the \author .. \affiliation  etc. as needed
% \email, \thanks, \homepage, \altaffiliation all apply to the current
% author. Explanatory text should go in the []'s, actual e-mail
% address or url should go in the {}'s for \email and \homepage.
% Please use the appropriate macro foreach each type of information

% \affiliation command applies to all authors since the last
% \affiliation command. The \affiliation command should follow the
% other information
% \affiliation can be followed by \email, \homepage, \thanks as well.
\author{Timo Reckling and Carsten Honerkamp}
\email[]{honerkamp@physik.rwth-aachen.de}
%\homepage[]{Your web page}
%\thanks{}
%\altaffiliation{}
\affiliation{
Institute for Theoretical Solid State Physics, RWTH Aachen University,
D-52056 Aachen and JARA - Fundamentals of Future Information
Technology}
%Collaboration name if desired (requires use of superscriptaddress
%option in \documentclass). \noaffiliation is required (may also be
%used with the \author command).
%\collaboration can be followed by \email, \homepage, \thanks as well.
%\collaboration{}
%\noaffiliation

\date{March 22, 2018}

\begin{abstract}
The functional renormalization group has become a widely used tool for the analysis of the leading low-temperature correlations in weakly to moderately coupled many-fermion lattice systems. A bottleneck for quantitatively more precise results is that the treatment of the frequency dependence of the flowing interactions is numerically quite demanding. Yet the frequency dependence is needed to compute relevant selfenergies and hence for controlled results on the energy scales for ordering or for the quasiparticle properties.  Here we explore an approximate parametrization of the frequency dependence of the interaction vertex that is inspired by established simplifications in the theory of superconductivity and that keeps the numerical effort bounded. We demonstrate the validity of the approximation for Cooper pairing problems and apply it to the two-dimensional Hubbard model.  
\end{abstract}

% insert suggested PACS numbers in braces on next line
% insert suggested keywords - APS authors don't need to do this
%\keywords{}

%\maketitle must follow title, authors, abstract, \pacs, and \keywords
\maketitle

% body of paper here - Use proper section commands
% References should be done using the \cite, \ref, and \label commands
\setlength{\parindent}{0pt}
\section{Introduction}
Functional Renormalization group (fRG) methods have been employed in a number of two-dimensional condensed matter systems as a tool to estimate the leading ordering tendencies and to derive phase diagrams (for reviews, see Refs. \onlinecite{metzner,platt}). They are especially useful in cases with competing orders, as they sum up large parts of the perturbation expansion in the interaction without particular bias toward any ordering channel. Yet, the bulk of the fRG work on one-band Hubbard models and virtually all multi-band systems has been done using a standard approximation that consists in neglecting the frequency dependence of the interactions and the selfenergy feedback.  Only a few works have included the frequency dependences\cite{karrasch,tam,honfulee,husemann12} and for two-dimensional models, more complete flows with frequency dependence and selfenergy feedback have only been performed in a few works\cite{giering,uebelacker,eberlein15,taranto,vilardi} with limited scope and only for one-band models. Both, for a leaner numerical treatment as well as for the treatment of multi-band or otherwise extended models, useful approximations for the frequency dependence are desirable. 

The main difficulty for the treatment of the frequency dependence arises from the fact that the two-particle  interaction vertex is a function of three Matsubara frequencies. This means that the memory consumption and the numerical effort scale at least with the third power of the number $N_\omega$ of captured Matsubara frequencies. This has to be multiplied with the similarly large effort for the description of the momentum dependence. Already some years back, simplified descriptions of the frequency  dependence were proposed\cite{karrasch,husemann12,giering}, leading effectively to one-frequency parametrizations. These descriptions bear a lot of resemblance to the frequency dependences known from single-channel approaches such as the ladder approximation or random phase approximation (RPA) and should hence contain meaningful physics. 
While for the main aspects, the qualitative results obtained by these approximations seem useful, in the case of the two-dimensional Hubbard model, the approximate treatment produced previously unknown dynamic instabilities in the charge channel\cite{husemann12} which were recently argued\cite{vilardi} to be artifacts of the approximation. In addition, studies of the strongly coupled situation\cite{rohringer12,kinza} made clear that the full interaction vertex contains structures that cannot be described by one-frequency parametrizations. So it is possibly unrealistic to expect that a single simplified treatment of the frequency dependence works for all purposes. Yet, one may also take the viewpoint that is better to incorporate some physically meaningful frequency dependence than to ignore it completely or only resolve important features at best crudely because of numerical limitations. 

In this work, we consider two numerically advantageous one-frequency parametrizations of the interaction vertex that can be regarded as further simplifications of the one-frequency schemes proposed earlier\cite{karrasch,husemann12}, partially taking up more recent developments in the field on inhomogeneous one-dimensional systems\cite{bauer,weidinger,markhof}. These modifications increase the effort at most quadratically in the number of frequencies (linearly without selfenergy feedback) compared to the flows without frequency dependence. They should allow one, in the next steps, to obtain meaningful selfenergies and should not be prone to develop spurious dynamical instabilities either. The main guiding example for the quality assessment of these approximations is the Cooper problem for phonon-mediated retarded interactions for whose frequency dependence various approximation levels exist. We demonstrate the qualitative and in many cases quantitative correctness of the approximations in a simplified pairing problem and in the 2D Hubbard model with and without phonon-mediated interactions.  

\section{Model and method}
\subsection{Model}
The Hubbard model is defined by the Hamiltonian 
\begin{equation}
\label{ }
H = -t \sum_{\langle ij \rangle,s } \left[ c_{i,s}^\dagger c_{j,s} +  c_{j,s}^\dagger c_{i,s} \right]
-t' \sum_{\langle\langle ij \rangle \rangle,s } \left[ c_{i,s}^\dagger c_{j,s} +  c_{j,s}^\dagger c_{i,s} \right]  + U \sum_{i} n_{i,\uparrow} n_{i,\downarrow} \, . 
\end{equation} 
Here, the sum in the first term is over the pairs of nearest neighbors $\langle ij \rangle$ and in the second term next-nearest neighbors $\langle\langle ij \rangle\rangle $ of the  lattice and spin projection $s=\uparrow$, $\downarrow$. In this paper we focus on the two-dimensional square lattice, but the approximate treatment of the frequency dependence  can certainly be used for other lattices as well. 
The parameter is $t$ is the hopping amplitude and $U$ measures the onsite repulsion between electrons with opposite spin projection. The first two terms define the non-interacting band structure $\epsilon (\vec{k}) = -2 t (\cos k_x + \cos k_y)- 4t' \cos k_x \cos k_y$, where we have set the lattice constant to unity.  We also add a Holstein-Einstein-phonon with frequency $\Omega_{\mathrm{ph}}$ to the Hamiltonian. It couples locally with a coupling strength $g$ to the electron density. This amounts to a term
\begin{equation}
\label{ }
H_{\mathrm{ph} } = \sum_i \Omega_{\mathrm{ph}}  b_i^\dagger b_i + g \sum_{i,s}   (b_i^\dagger + b_i ) \, n_{i,s} \, . 
\end{equation}
In the functional integral formalism, the fermions are represented by Grassmann numbers and the phonons by complex fields. As the phonons are non-interacting the action is quadratic in the phonon fields and one can integrate them out. This gives rise to a phonon-mediated electron-electron interaction
 \begin{equation}
\label{phonmed}
S_{\mathrm{ph}} = \frac{T}{2V} \sum_{k,k',q \atop s,s'} V_{\mathrm{ph}} (q) \bar{c}_{k+q,s}Ê\bar{c}_{k'-q,s'} c_{k',s'} c_{k,s} \, ,Ê
\end{equation}
with the phonon-mediated attraction
\begin{equation}
\label{vph}
V_{\mathrm{ph}} (q) = - \frac{2 g^2\Omega_{\mathrm{ph}}}{q_0^2 + \Omega_{\mathrm{ph}}^2}Ê= 
- V_{\mathrm{ph},0} \frac{\Omega_{\mathrm{ph}}^2}{q_0^2 + \Omega_{\mathrm{ph}}^2} \, .
\end{equation}
Additional model parameters are the chemical potential $\mu$, which tunes the band filling, and the temperature $T$.

\subsection{fRG equation for the two-particle vertex} 
Functional renormalization group methods have been applied to the Hubbard-like fermion lattice  model in different forms and with different degree of approximations\cite{metzner,platt}. The standard renormalization group formalism based on the Wetterich equation\cite{wetterich} prescribes
flow equations as function of a flowing energy scale, here called $\Lambda$, for the one-particle irreducible vertex functions of the theory. In this work we focus on the two-particle interaction vertex and ignore all higher vertices as well as the flow of the one-particle vertex, i.e. the selfenergy.  

\begin{figure}
 \includegraphics[width=.48\columnwidth]{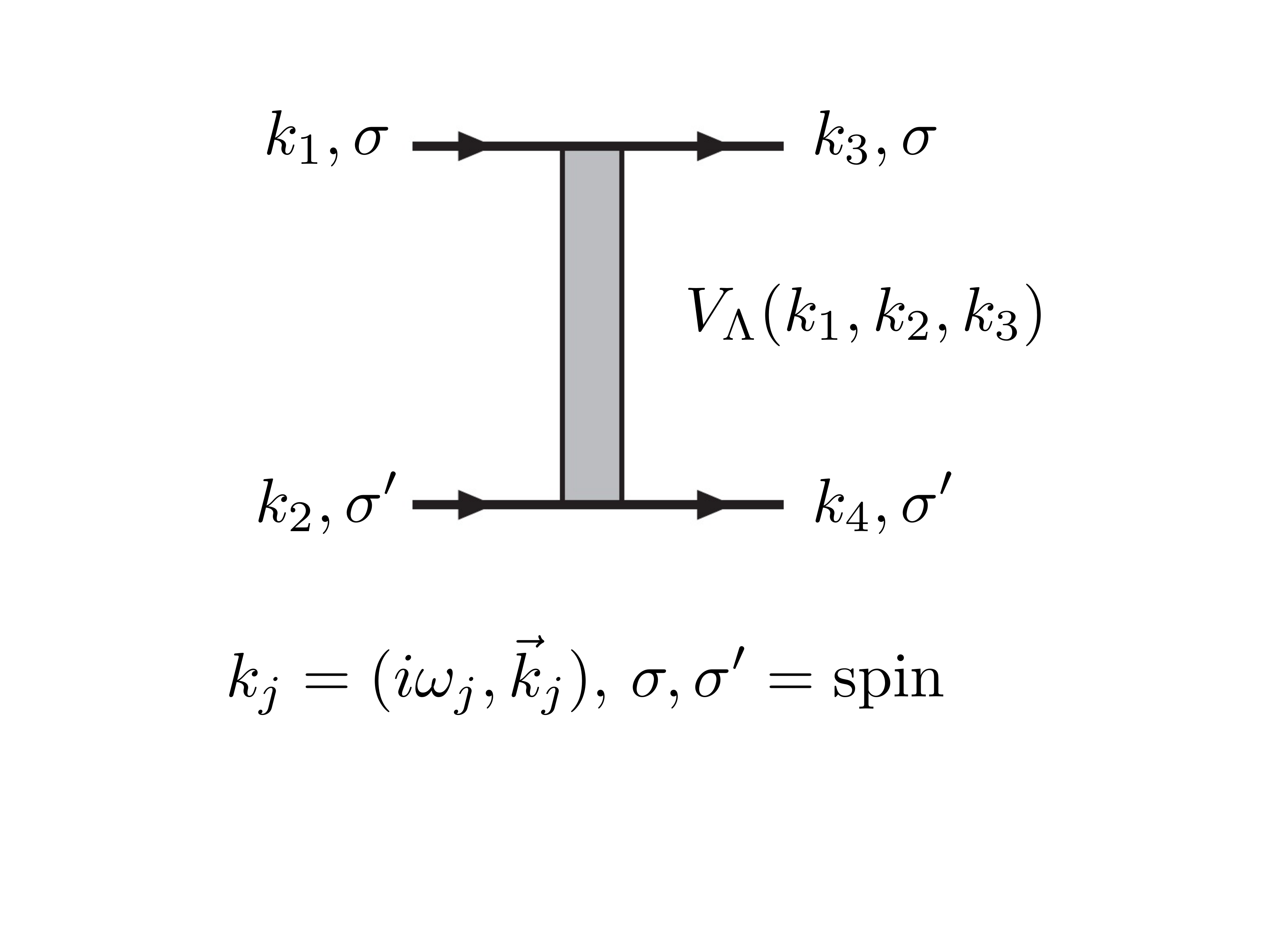}
 \caption{The coupling function $V_\Lambda (k_1,k_2,k_3)$ from which the full spin-dependent one-particle irreducible vertex can be reconstructed in the case of spin-rotational invariance.}
  \label{Vfig}
\end{figure}

\begin{figure}
 \includegraphics[width=.79\columnwidth]{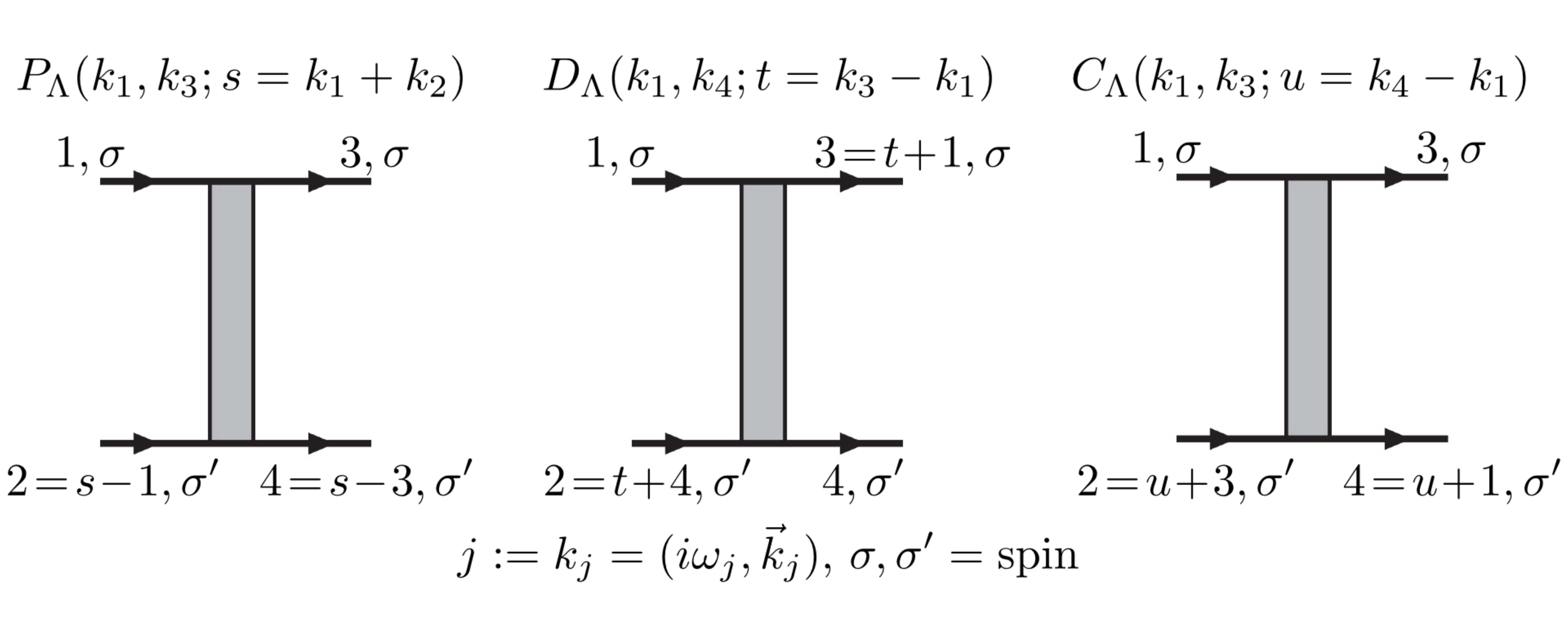}%
 \caption{The three channel couplings $P_\Lambda (k_1,k_3;s=k_1+k_2)$, $D_\Lambda (k_1,k_4;t=k_3-k_1) $ and $C_\Lambda (k_1,k_3; u = k_4-k_1) $ used for the decomposition of the full interaction in Eq. \ref{Vdecomp}.}
  \label{PDCfig}
\end{figure}

Let us first work with combined Matsubara-frequency/wavevector variables $k_i=(k_{0,i},\vec{k}_i)$ where a fermionic Matsubara frequency $k_{0,i}$ is an odd multiple of $\pi T$ and $\vec{k}_i$ is a wavevector in the first Brillouin zone. In a one-band model with spin-rotational SU(2) invariance the two-particle vertex can be described by a coupling function $V_\Lambda (k_1,k_2,k_3)$. In this notation\cite{hsfr} (see also Fig. \ref{Vfig}), the two incoming particles $k_1$ and $k_2$ of the interaction carry spin projection $s$ and $s'$ and $k_3$ denotes the first outgoing particle with the same spin projection as $k_1$, i.e. $s$. 
The two-particle scattering can also be described with Mandelstam variables 
\begin{equation}
\label{ }
s= k_1+k_2 \, , \quad t=k_3-k_1 \, , \quad \mbox{and} \quad u= k_4-k_1 \, , 
\end{equation}
with $k_4=k_1+k_2-k_3$ (modulo lattice). The coupling function is then written as a sum of the bare non-retarded interaction $U$ and three channel couplings,
\begin{equation}
\label{Vdecomp}
V_\Lambda (k_1,k_2,k_3) = U + P_\Lambda (k_1,k_3;s) +  D_\Lambda (k_1,k_4;t)  
+ C _\Lambda (k_1,k_3;u)  \, .
\end{equation}
For a diagrammatic representation of these channel couplings see Fig.~\ref{PDCfig}.
The initial conditions for the channel couplings we choose 
\begin{equation}
\label{ }
P_{\Lambda_0} (k_1,k_3;s)  = 0 \, , \quad C_{\Lambda_0} (k_1,k_3;u)  = 0 \, , \quad \mbox{and} \quad
D_{\Lambda_0} (k_1,k_4;t)  = V_{\mathrm{ph}} (t) \,  ,
\end{equation}
using the the phonon-mediated interaction from Eq. \ref{vph}.

The flow equation in the common level-2 truncation\cite{metzner} for 
$V_\Lambda (k_1,k_2,k_3)$ reads
\begin{eqnarray}
\frac{d}{d\Lambda} V_\Lambda (k_1,k_2,k_3) =    \partial_\Lambda {P}_\Lambda (k_1,k_3;s)  + \partial_\Lambda {D}_\Lambda (k_1,k_4;t) + \partial_\Lambda {C}_\Lambda (k_1,k_3;u) 
\label{vdot} \end{eqnarray} 
with the one-loop particle-particle contributions $ \partial_\Lambda {P}_\Lambda (k_1,k_3;s)$ and the two different particle-hole channels $\partial_\Lambda {D}_\Lambda (k_1,k_4;t)$ and $\partial_\Lambda {C}_\Lambda (k_1,k_3;u) $, where
\begin{eqnarray} \label{pdot} \partial_\Lambda
{P}_\Lambda (k_1,k_3;s)  & = & \frac{T}{N_L} \sum_k V_\Lambda (k_1,-k_1+s,k) \, 
\partial_\Lambda \left[ {G}_{\Lambda} (k) {G}_{\Lambda} (-k+s) \right] V_\Lambda (k,-k+s,k_3) 
 \\ \label{ddot} 
\partial_\Lambda {D}_\Lambda (k_1,k_4;s)  & = & -2 \frac{T}{N_L} \sum_k V_\Lambda (k_1,k+t,k_1+t) \, 
\partial_\Lambda \left[ {G}_{\Lambda} (k) {G}_{\Lambda} (k+t) \right] V_\Lambda (k,k_4+t,k+t) \\ \nonumber && +
\frac{T}{N_L} \sum_k V_\Lambda (k_1,k+t,k_1+t)  \, 
\partial_\Lambda \left[ {G}_{\Lambda} (k) {G}_{\Lambda} (k+t) \right] V_\Lambda (k_4+t,k,k+t) \\ \nonumber  &&  +
\frac{T}{N_L} \sum_k V_\Lambda (k+t,k_1,k_1+t) \, 
\partial_\Lambda \left[ {G}_{\Lambda} (k) {G}_{\Lambda} (k+t) \right] V_\Lambda (k,k_4+t,k+t)
 \\ \label{cdot}
 \partial_\Lambda {C}_\Lambda (k_1,k_3;u)  & = & \frac{T}{N_L} \sum_k V_\Lambda (k_1,k+u,k) \, 
\partial_\Lambda \left[ {G}_{\Lambda} (k) {G}_{\Lambda} (k+u) \right] V_\Lambda (k,k_3+u,k_3) 
\end{eqnarray}
In these equations, the product of the two internal lines in the
one-loop diagrams contains the full single-particle Green's function  $G_{\Lambda} (k) = R_\Lambda (k) / \left[ - i \omega + \epsilon(\vec{k}) + R_\Lambda (k) \Sigma_\Lambda (k) \right]$ at RG scale $\Lambda$.   
The momentum- or energy-shell cutoff function or regulator ${R}_\Lambda (k) $ used in this work for all practical calculations suppresses the modes with $|\epsilon (\vec{k}) | < \Lambda$. Its derivative $\dot {R}_\Lambda (k)$ confines the modes  to a momentum shell with $|\epsilon (\vec{k}) | \approx \Lambda$. $N_L$ is the number of unit cells in the system which should be send to infinity to convert the momentum sums to integrals. 
Along with the flow of the vertex functions, we can also obtain the flow of susceptibilities in channels of interest. This is explained e.g. in the review Ref. \onlinecite{metzner} and in appendix of  Ref. \onlinecite{hsfr}.

\subsection{Approximations and treatment of the frequency dependence}
\begin{figure}
 \includegraphics[width=.79\columnwidth]{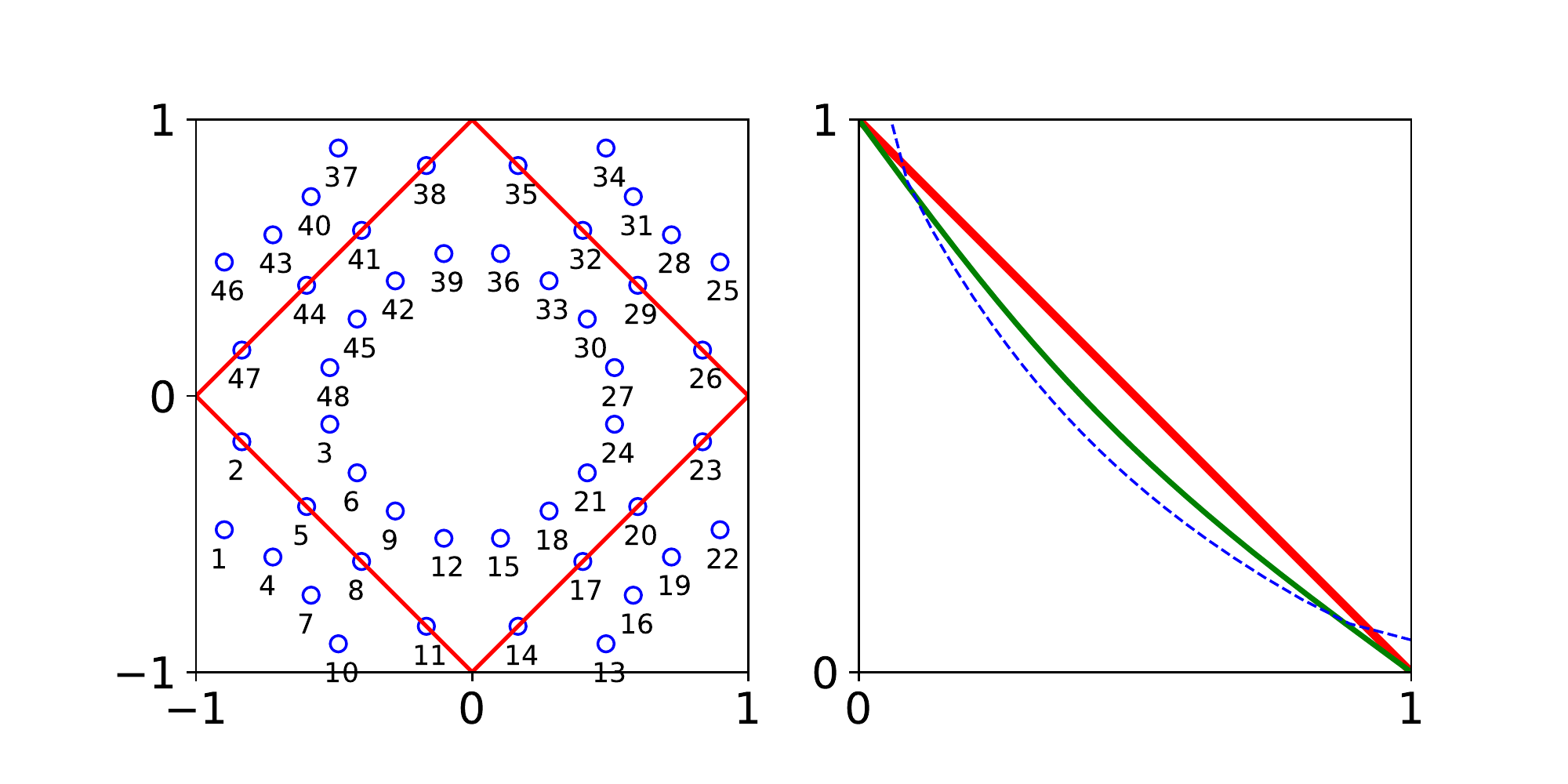}%
 \caption{Left: 48 points in the Brillouin zone used for discretizing the momentum dependence of the interaction vertex for the half-filled band with $t'=0$. The points are grouped in three rings with 16 points each. The central ring sits at the Fermi level, and the other two rings below and above the Fermi level at energy $|\epsilon | =1.8t$. Right: Exemplary Fermi surfaces in the first Brillouin zone quadrant for $\mu=0$, $t'=0$ (thick red line), $\mu=-0.6t$, $t'=-0.15t$ (thinner green line)  and $\mu=-0.95t$, $t'=-0.25t$ (dashed blue line). }
  \label{Fig1points}
\end{figure}

There are four main approximations involved in application of equations (\ref{vdot}), (\ref{pdot}), (\ref{ddot}) and (\ref{cdot}) in the present work. This paper mainly deals with ways to describe the frequency dependence of the interactions that is either present already in the bare interactions, e.g. if phonon-mediated interactions are considered, or that is generated by the loop corrections during the flow. Before we get to the treatment of the frequency dependence, we briefly discuss the other approximations:
\begin{itemize} 
\item We work in the level-2 truncation\cite{metzner} of the hierarchy of flow equations for the 1PI vertices that consists in dropping all vertices higher than the two-particle vertex. This confines the scheme to weak to moderate interactions, at least as quantitative results are concerned. Note however that recently extended fRG schemes have been devised\cite{eberlein14,kugler1,kugler2} that allow one to go beyond this truncation level. 
\item In this work we neglect the selfenergy dependence, as we are basically analyzing schemes that may allow us, in the next step, to compute the selfenergy in a more efficient way. 
\item For the wavevector dependence we use a Brillouin zone patching with 16 points on the Fermi surface and  another 16 points each above and below the Fermi surface (at $|\epsilon (\vec{k})| = \pm1.8 t$), essentially in the same way as explained in Ref. \onlinecite{honedhd}. For the location of the patch points, see also Fig. \ref{Fig1points}.
The neighborhoods of these 48 points define patches in which the coupling function is kept constant when one of the legs $\vec{k}_i$ moves inside a patch.  The argument for the choice of the $N$-patch scheme is mainly its simplicity and the possibilities to compare with many previous works.
\end{itemize} 
The quantitative changes by the inclusion of the selfenergy and multi-loop effects  in the Hubbard model will described in future work. Next we discuss three ways to capture the frequency dependence.

\subsubsection{Three fermionic Matsubara frequencies}
In the most direct, 'brute-force' approach we use $N_\omega$  fermionic Matsubara frequencies $-(N_\omega-1) \pi T$, $-(N_\omega-3) \pi T$, $\dots$, $(N_\omega-1)  \pi T$ for each of the three frequency variables in $k_1$, $k_2$ and $k_3$ in the coupling function $V_\Lambda (k_1,k_2,k_3)$. In the internal loop summations that are extended beyond these frequencies, the coupling function is also needed at higher frequencies. In such cases the boundary values at $\pm (N_\omega-1)  \pi T$ of the coupling function are used. We have done runs with $N_\omega=6,8$ and 10  Matsubara frequencies, with little quantitative changes. Note that other recent fRG studies emphasize the importance of the frequency dependence and selfenergy feedback\cite{eberlein15,vilardi}. There are also more advanced techniques to treat the high frequency tails outside the frequency considered explicitly\cite{wentzell,vilardi,tagliavini}.

The direct description with three Matsubara frequencies has been performed in various forms in recent literature\cite{honfulee,tam,uebelacker,vilardi}. Most likely it should be considered the best choice because it involves the least approximations.  There are also advanced prescriptions for how to deal best with the large frequencies outside the window given by $N_\omega$\cite{wentzell,vilardi,tagliavini}. However, from the numerical perspective, using three frequencies considerably increases the effort compared to the case without frequency dependence by two factors. One is that the vertex function size goes up by a  factor  $N_\omega^3$, leading to larger memory requirements and also more complicated data evaluations.  Second, the runtime scales at least $\sim N_\omega^4$, as there is an additional internal summation on the right hand side of the flow equations that cannot be done analytically anymore with general frequency-dependent interactions. Note that ongoing studies\cite{tagliahille} and this work suggest that at not too low temperatures it is possible to get away with a handful Matsubara frequencies such that the overall numerical effort is still bearable. Nevertheless, a speed-up is desirable. especially regarding applications in multiband or three-dimensional models.

\subsubsection{Parametrization with one frequency per channel: static channel coupling} 
Recent fRG approaches\cite{husemann,wang,lichtenstein} have explored the use of channel decompositions into three different interaction channels, based on the observation that the effective vertices consist of structures that can be described by varying a single Mandelstam variable, i.e. a momentum or frequency transfer or the total momentum or frequency. We again write the full coupling function as in Eq. \ref{Vdecomp}.
%\begin{equation}
%\label{PDCdecomp}
%V_\Lambda (k_1,k_2,k_3) = U + P_\Lambda (k_1,k_3;s= k_1+k_2) + D_\Lambda (k_1,k_4;t= k_3-k_1) + %C_\Lambda (k_1,k_3;u= k_4-k_1) \, .
%\end{equation}
The channel couplings $P_\Lambda$, $C_\Lambda$, $D_\Lambda$ are then expanded in a form factor expansion of the type
\begin{equation}
\label{Pexp}
P( k_1,k_3;s) = \sum_{l,l'} P_{ll'} (s) f_l (k_1) f^*_l (k_3) \, , 
\end{equation}
with form factors $f_l(k)$ that form a complete basis in the space in which $k$ lives. Regarding the momentum components, it has been established that the main contributions arise from $f_l$s that vary mildly if $\vec{k}$ moves in the Brillouin zone\cite{lichtenstein}. Hence the expansion in Eq. \ref{Pexp} can usually be well approximated by the first few terms. In real space this corresponds to a interaction channel in which bilinears of a particular type with rather small extension interact. The stronger the $s$-dependence of $P_{ll'} (s)$ is, the more long-ranged the interaction between the short-ranged bilinears becomes. 

Regarding the frequency dependence, it is a priori unclear if a truncated expansion is a good approximation and which form factors or basis function should be chosen. The simplest approach, leaving this second issue mainly untouched, may be to truncate the expansion in the frequency basis already after the first term that is chosen to be just constant. Note that this approximation is exact in second order perturbation theory for the Hubbard interaction where the contributions  only depend on either $s$, $t$ or $u$. In fact, working with channel couplings $P_\Lambda$, $C_\Lambda$ and $D_\Lambda$ that only depend on one frequency was shown to be a reasonable approximation in zero-dimensional\cite{karrasch} and one-dimensional models\cite{bauer,weidinger,markhof}. In two dimensions, a one-frequency parametrization (with adaptive boson-fermion vertices) was used by Husemann et al.\cite{husemann12}, but discovered intriguing divergences at non-zero Matsubara frequencies in the density channel.  These were argued to be spurious instabilities that do not appear in the three-frequency descriptions by Vilardi et al.\cite{vilardi}. Here we propose to retain the one-frequency parametrization, however in an even more simplified scheme that should also be immune against any spurious finite-frequency divergences.  

To be more concrete, we proceed to write the coupling function as 
\begin{equation}
\label{VPDC}
V_\Lambda (k_1,k_2,k_3) = U + P_\Lambda (\vec{k}_1,\vec{k}_2,\vec{k}_3; s_0)  + C_\Lambda (\vec{k}_1,\vec{k}_2,\vec{k}_3; u_0) + D_\Lambda (\vec{k}_1,\vec{k}_2,\vec{k}_3; t_0 ) \, .
\end{equation}
Here we employ the $N$-patch description of the wavevector dependence on $\vec{k}_1,\vec{k}_2,\vec{k}_3$. The main reason for this is that we want to compare the results closely with the $N$-patch codes without frequency dependence  (as a naive standard) and with the three-frequency parametrization as described above. In the spirit of a formfactor expansion of the frequency dependence that is truncated after the first term, we allow the pairing channel $P_\Lambda$ to depend on the total frequency $s_0$ and the direct and crossed particle-hole channels, $D_\Lambda$ and $C_\Lambda$ on the respective transfer frequencies $t_0$ and $u_0$. When we now insert the decomposition (\ref{VPDC}) into the right hand side of the flow equations we get, e.g. for the particle-particle (PP) diagrams,
\begin{eqnarray}
\partial_\Lambda
{P}_\Lambda (k_1,k_3;s) &= &  \frac{T}{N_L} \sum_{\vec{k}, i \omega} \, 
\left[ U + P_\Lambda ( \vec{k}_1,\vec{k}_2,\vec{k}; s_0 ) + D_\Lambda (\vec{k}_1,\vec{k}_2,\vec{k} ; t_0 = \omega - k_{0,1} ) + C_\Lambda (\vec{k}_1,\vec{k}_2,\vec{k} ; u_0 = \omega - k_{0,2}) \right]
\nonumber\\
&&   \, \cdot  \, \partial_\Lambda \left[ G_\Lambda( \vec{k}, \omega) G_\Lambda( -\vec{k}+ \vec{k}_1+\vec{k}_2, - \omega+s_0) \right] \cdot
\nonumber\\
&&  
\left[ U + P_\Lambda ( \vec{k},-\vec{k}+ \vec{k}_1+\vec{k}_2 ,\vec{k}_3; s_0 ) + D_\Lambda (\vec{k},-\vec{k}+ \vec{k}_1+\vec{k}_2 ,\vec{k}_3 ; t_0 = \omega - k_{0,3} ) \right. \nonumber\\
&&   \left. + C_\Lambda (\vec{k},-\vec{k}+ \vec{k}_1+\vec{k}_2 ,\vec{k}_3  ; u_0= \omega - k_{0,4} ) \right] \label{PPdia}
\end{eqnarray}
The equations for the other channel couplings follow analogously. 
We observe that for the PP channel the summation frequency $i \omega$ appears in the frequency arguments of the 'channel-non-native' particle-hole-channel couplings $C_\Lambda$ and $D_\Lambda$, while it does not appear in the argument in the channel-native coupling $P$. Likewise, in the direct particle-hole  channel, the summation frequency will appear in the channel-non-native $P_\Lambda$ and $C_\Lambda$ but not in the channel-native coupling $D_\Lambda$. This appearance of the summation frequency in the arguments thwarts the independent and fast calculation of the bubble sums. A simple approximation to avoid this problem, also used by Bauer et al.\cite{bauer}, Weidinger et al.\cite{weidinger}, and  Markhof et al.\cite{markhof} under the name coupled-ladder approximation for inhomogeneous one-dimensional systems, is the {\bf static channel-coupling approximation}, where one inserts the zero-frequency values for the non-native couplings. This gives
 \begin{eqnarray}
\partial_\Lambda
{P}_\Lambda (k_1,k_3;s) &=&  \frac{1}{N_L} \sum_{\vec{k}} \, 
\left[ U + P_\Lambda ( \vec{k}_1,\vec{k}_2,\vec{k}; s_0 ) + D_\Lambda (\vec{k}_1,\vec{k}_2,\vec{k} ; t_0 = 0 ) + C_\Lambda (\vec{k}_1,\vec{k}_2,\vec{k} ; u_0 = 0 ) \right]
\nonumber\\
&&   \, \cdot T \sum_{i\omega}  \partial_\Lambda\left[ {G} ( \vec{k}, \omega) G( -\vec{k}+ \vec{k}_1+\vec{k}_2, - \omega+s_0)  \right] \cdot
\nonumber\\
&&  
\left[ U + P_\Lambda ( \vec{k},-\vec{k}+ \vec{k}_1+\vec{k}_2 ,\vec{k}_3; s_0 ) + D_\Lambda (\vec{k},-\vec{k}+ \vec{k}_1+\vec{k}_2 ,\vec{k}_3 ; t_0 =0 ) + C_\Lambda (\vec{k},-\vec{k}+ \vec{k}_1+\vec{k}_2 ,\vec{k}_3  ; u_0= 0 ) \right]\nonumber\\
&&   \label{PPdiaD}
\end{eqnarray}
This way, the couplings can be pulled out of the frequency sums. This results in a major numerical speedup, as now the internal Matsubara sums can be performed separately  without calling up vertices. 

In passing we can compare the static channel-coupling approximation to the fully static flows without any frequency dependence, in which all one-loop diagrams on the right hand side are evaluated with zero transfer frequency or total frequency, respectively. Now, in the static channel-coupling approximation, only the zero-frequency components of the channels couple into each other. Furthermore, within one channel, the different bosonic frequency components do not couple either. The equations in one channel, say for $P_\Lambda (\vec{k}_1,\vec{k}_2,\vec{k}_3;s_0)$,  are basically RPA-like, i.e. quadratic in $P_\Lambda$ with the same $s_0$, and with additional contributions that are linear in $P_\Lambda$ at this $s_0$, involving a frequency-independent factor that only contains the $t_0=0$ and $u_0=0$ $D_\Lambda$ or $C_\Lambda$-couplings. Hence, the flow for the zero-frequency components is the same as in the fully static scheme and the critical scales of static channel coupling and of fully static fRG coincide. We also confirmed this finding numerically. Furthermore, for frequency-independent initial interactions, the evolving frequency-dependence of the couplings is fully determined by the one-loop diagrams that are monotonous in their bosonic frequency. Therefore, strange frequency dependences with peaks developing at nonzero frequencies should not occur in this approximation.

\subsubsection{Shortcomings of the static channel-coupling approximation}Ê
It is obvious that the static channel-coupling approximation includes important physical mechanisms like mutual screening or amplification of the different interaction channels and thus goes beyond simple random phase or ladder approximations. However, there are well-known issues that are not captured by this approximation. Consider the Cooper pairing problem of electrons due to a phonon-mediated interaction of the type (\ref{phonmed}). In the channel-decomposed fRG scheme, the Cooper problem corresponds to studying the P-channel and neglecting the flows in the C- and D-channel. The phonon-mediated interaction (\ref{phonmed}) is naturally included as initial condition for the D-channel which appears then inserted in the PP diagram on the right hand side of the P-flow equation (\ref{PPdiaD}). Here, if we imagine small external frequencies, the finite width $\Omega_{\mathrm{ph}}$ of the phonon-propagator effectively suppresses the contributions of higher summations frequencies. An often-used approximation is to impose a high-frequency cutoff $|\omega | < \Omega_{\mathrm{ph}}$ on these summations. 
Then, it has also become customary in textbook treatments of superconductivity\cite{agd} to replace the frequency cutoff with an energy cutoff that restricts the electronic dispersion $|\epsilon (\vec{k})|$ to values smaller than  $\Omega_{\mathrm{ph}}$, instead of the original restriction that $|\epsilon (\vec{k})|$ is smaller than the electronic band width $W$. At least for simplified situations with $\Omega_{\mathrm{ph}} \ll W$, this gives the same answer for the critical temperature $T_c$ or the energy gap $\Delta$. The upshot of the arguments is that the critical temperature has now $\Omega_{\mathrm{ph}}$ as prefactor,
\begin{equation}
\label{ }
T_c = 1.13 \, \Omega_{\mathrm{ph}} e^{-1/\rho_0 V_{\mathrm{ph}}} \, ,
\end{equation} 
with the (on scale $\Omega_{\mathrm{ph}}$ assumed constant) density of states $\rho_0$. If we had ignored the finite frequency-width of the phonon propagator we would have obtained $T_c = 1.13 \, W  e^{-1/\rho_0 V_{\mathrm{ph}}}$. This would also be the result of the static channel-coupling approximation. Hence, at least for stronger frequency dependence in the initial condition, the static approximation for the inserted non-native channels (as D in the PP channel) is insufficient.

\subsubsection{At-scale channel-coupling approximation}
Hence the question arises how the finite width of non-native channels inserted on the right hand side of the flow equations can be resolved. A realistic workaround unfolds if one considers a Matsubara frequency cutoff RG flow. For such a flow, at least the internal propagator line that carries a differentiated cutoff function $\dot{R}_\Lambda$ has a Matsubara frequency $\omega$ at the running RG scale $\Lambda$. The other line has a frequency whose absolute values is equal or higher.
If we again consider small external frequencies, the frequency transfer e.g. in the PP channel will be at least of order $\omega \approx \Lambda$. Hence we should use the coupling at frequency transfer $\Lambda$ instead of the one at frequency transfer zero.  For the PP channel, this gives
 \begin{eqnarray}
\lefteqn{  \partial_\Lambda {P}_{\Lambda} (k_1,k_3;s_0) = }\nonumber\\
&&  \frac{1}{N_L} \sum_{\vec{k}} \, 
\left[ U + P_\Lambda ( \vec{k}_1,\vec{k}_2,\vec{k}; s_0 ) + D_\Lambda (\vec{k}_1,\vec{k}_2,\vec{k} ; t_0 = \Lambda ) + C_\Lambda (\vec{k}_1,\vec{k}_2,\vec{k} ; u_0 = \Lambda ) \right]
\nonumber\\
&&   \, \cdot T \sum_{i\omega} \dot{G}_\LambdaÊ( \vec{k}, \omega) G_\Lambda( -\vec{k}+ \vec{k}_1+\vec{k}_2, - \omega+s_0) \cdot
\nonumber\\
&&  
\left[ U + P_\Lambda ( \vec{k},-\vec{k}+ \vec{k}_1+\vec{k}_2 ,\vec{k}_3; s_0 ) + D_\Lambda (\vec{k},-\vec{k}+ \vec{k}_1+\vec{k}_2 ,\vec{k}_3 ; t_0 = \Lambda ) + C_\Lambda (\vec{k},-\vec{k}+ \vec{k}_1+\vec{k}_2 ,\vec{k}_3  ; u_0= \Lambda ) \right] \label{PPdiaE} \, .
\end{eqnarray}
Just like the static approximation, this {\bf at-scale channel-coupling} scheme has the advantage that the couplings are pulled out of the Matsubara sums. If we apply the at-scale channel coupling to the phonon-mediated Cooper problem described above, it basically means that the $D$-coupling only starts to influence the P-coupling if $\Lambda \le \Omega_{\mathrm{ph}}$. This has again the effect of reducing the critical temperature or scale to be $\sim \Omega_{\mathrm{ph}}$. Hence, while keeping the simplicity of the static coupling, the at-scale channel-coupling approximation also implements retardation effects correctly. Again, considering now a band energy cutoff, we can go one step further and relate the band energies (instead of the frequency) that are now confined  to $|\epsilon (\vec{k})| \ge \Lambda $  in the loops with the frequency of the non-native couplings.\footnote{That this works can be seen by analyzing e.g. the particle-particle diagram at zero total momentum and frequency, which basically gives with a sharp frequency cutoff (taking a constant density of states $\rho_0$ and bandwidth $W \gg \Lambda$) 
\begin{equation}
\label{ }
\int_{-W}^W d\epsilon \, \rho_0 \int \frac{d\omega}{2\pi}  \dot{G}_\LambdaÊ( \vec{k}, \omega) G_\Lambda( -\vec{k}, - \omega)  =   \frac{1}{2\pi}Ê\int_{-W}^W  d\epsilon \, \rho_0
\frac{1}{\Lambda^2  + \epsilon^2} \approx  \frac{\rho_0}{2 \Lambda}  \, . 
\end{equation} 
We would get the same for a energy-shell cutoff, leading to 
\begin{equation}
\label{ }
\int_{-W}^W d\epsilon \, \rho_0 \int \frac{d\omega}{2\pi}  \dot{G}_\LambdaÊ( \vec{k}, \omega) G_\Lambda( -\vec{k}, - \omega)  =   \frac{1}{2\pi}Ê\int d\omega \, \rho_0
\frac{1}{ \omega^2 + \Lambda^2  } =  \frac{\rho_0}{2 \Lambda}  \, . 
\end{equation} 
This equivalence argument appears to be quite crude as it is based on comparing specific situations. We will however see below that the at-scale approximation applied in the energy-shell flow gives useful results.} 

We can also reconstruct the three-frequency dependence in the one-channel approximations by the formula
\begin{equation}
\label{VPDC}
V_\Lambda (k_1,k_2,k_3) = U + P_\Lambda (\vec{k}_1,\vec{k}_2,\vec{k}_3; s_0 =k_{0,1}+k_{0,2} ) + D_\Lambda (\vec{k}_1,\vec{k}_2,\vec{k}_3; t_0 = k_{0,3}- k_{0,1}  ) + C_\Lambda (\vec{k}_1,\vec{k}_2,\vec{k}_3; u_0=k_{0,3}-k_{0,2}  ) \, .
\end{equation}
This will be used in Figs. \ref{CooperFreq}, \ref{CooperUFreq} and \ref{UvhFreq} where we compare the frequency dependence of the three-frequency parametrization with that of the at-scale one-frequency approximation.

\section{Numerical evaluation}
Here we test the two channel-decomposed one-frequency approximations against the 'full' flow with a vertex that depends on three fermionic Matsubara frequencies.  

\subsection{Phonon-mediated Cooper instability}  
\begin{figure}
 \includegraphics[width=.79\columnwidth]{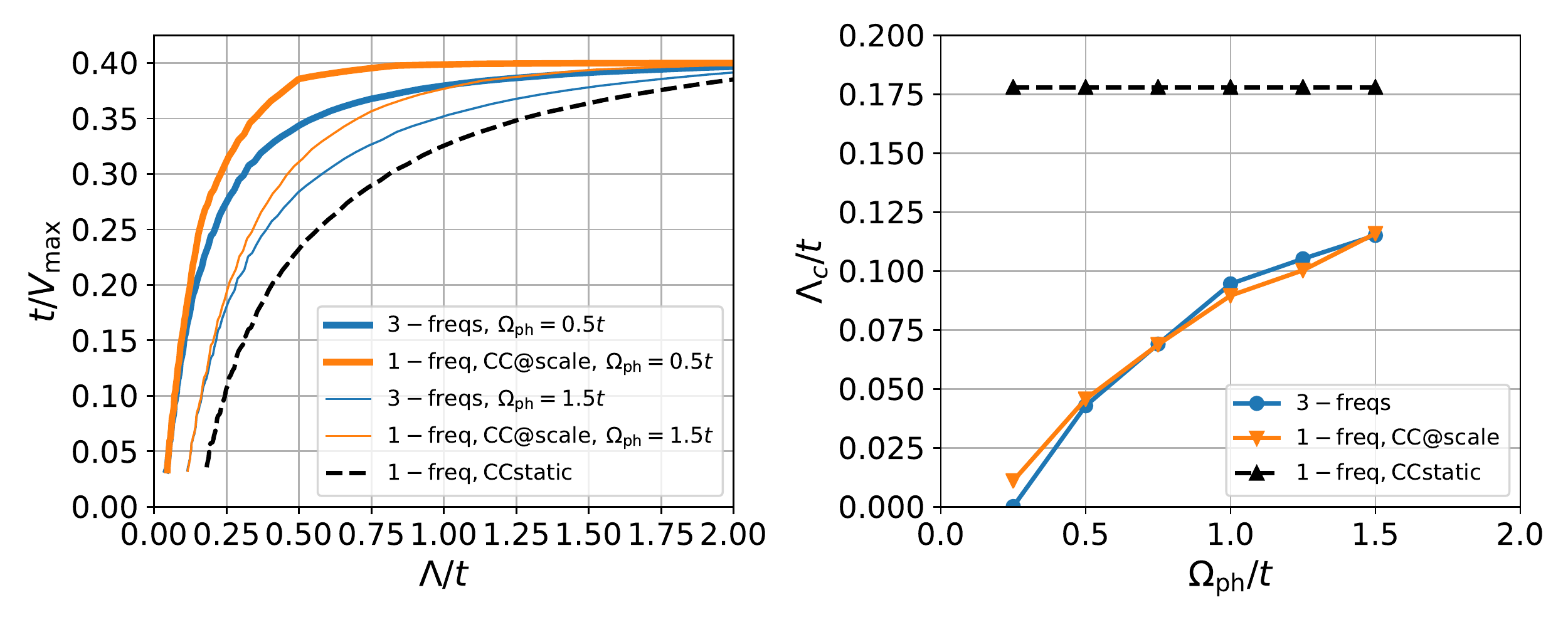}%
 \caption{Flows to strong coupling in particle-particle channel only, with phonon-mediated bare interaction. Left plot: RG flows of the inverse maximal coupling constants for two different phonon frequencies in three different treatments of the frequency dependence, three fermionic Matsubara frequencies without channel decomposition, and the two one-frequency parametrizations with static and at-scale channel-coupling. The latter curve start to dives to when the cutoff gets below the phonon frequency.
 The static channel-coupling results are independent of the phonon frequency $\Omega$, as only the peak of the phonon-propagator at zero frequency enters in the particle-particle channel.  Right plot: Critical RG scale as function of the phonon frequency for the three schemes. All data at $T=0.05t$, $\mu=-0.95t$ and $t'=-0.25t$. }
  \label{Cooper}
\end{figure}

First, as a sanity check with compare the three-frequency parametrization with the two one-frequency schemes for a pure Cooper instability without any direct electron-electron interactions. The initial condition for the flow is given by Eq. \ref{phonmed}). Hence we have initially 
\begin{equation}
\label{ }
D_{\Lambda_0} (k_1,k_4;t) =  - V_{\mathrm{ph},0} \frac{\Omega_{\mathrm{ph}}^2}{t_0^2 + \Omega_{\mathrm{ph}}^2} \end{equation}
as well as 
\begin{equation}
P_{\Lambda_0} (k_1,k_3;s) = 0 \, , \quad 
C_{\Lambda_0} (k_1,k_3;u) = 0 \, , \quad 
U= 0 \, .
\end{equation}
For this setting we run the flow in the Cooper channel only, i.e. only $\partial_\Lambda P_\Lambda (k_1,k_3;s) $ is considered and all particle-hole terms on the right hand side of the flow equation are dropped. The attractive interaction flows to strong coupling in all three schemes. In Fig. \ref{Cooper} we show the flows of the maximal (absolute value) component of the interaction for the three different schemes, for two different phonon frequencies. One can clearly see that the three-frequency scheme and the one-frequency scheme with at-scale channel coupling  diverge at quite comparable scales. In the latter scheme, the flow only really starts at scales $\Lambda \approx  \Omega_{\mathrm{ph}}$, as it should be based on its construction. As expected, the one-frequency scheme with static channel coupling does not yield a good approximation of the instability scale.
In the right plot we show the critical scales for the Cooper instability as function  of the phonon frequency $\Omega_{\mathrm{ph}}$, again showing that the one-frequency scheme with at-scale channel coupling approximates the instability scales of the more precise three-frequency flow quite well. In contrast with this, the static channel-coupling scheme does not reflect the dependence of the instability scale on $\Omega_{\mathrm{ph}}$. 

\begin{figure}
 \includegraphics[width=.79\columnwidth]{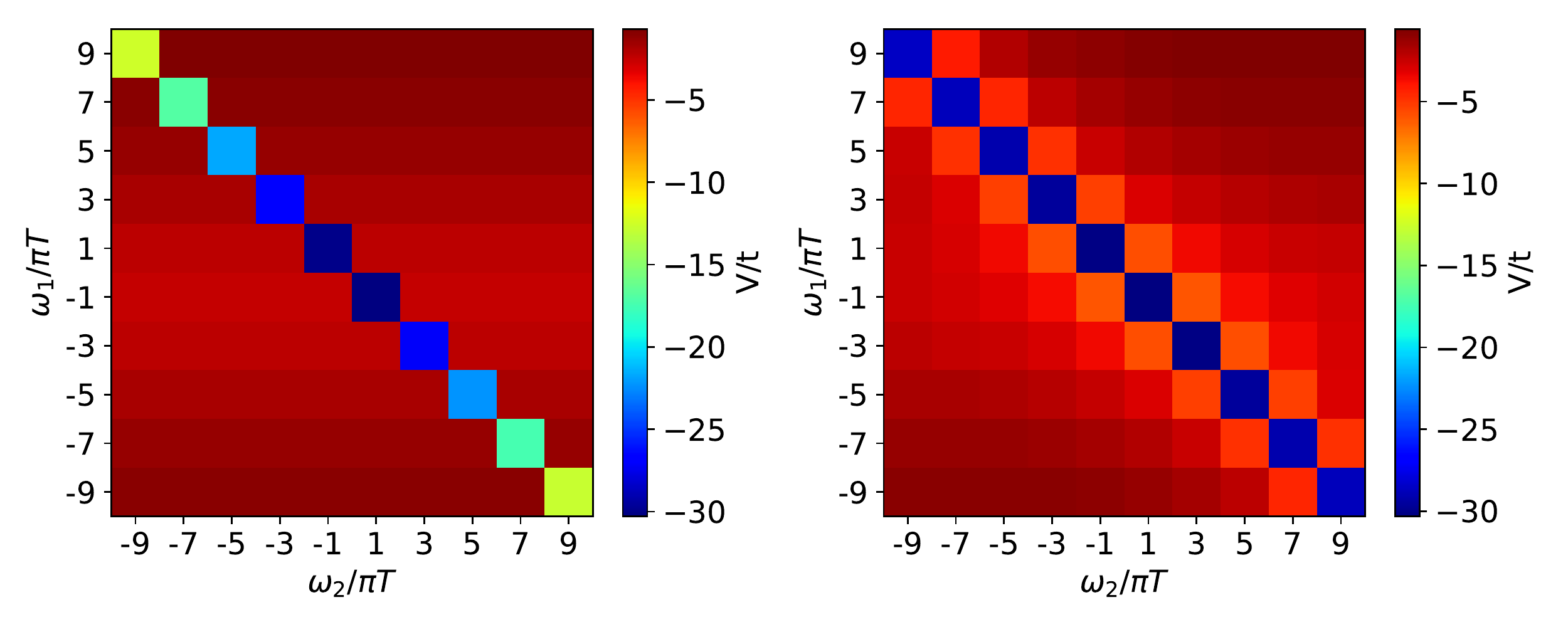}%
 \caption{Frequency dependence of the diverging pairing interaction. The colorbar denotes the strength of the interaction. $\omega_1$ and $\omega_2$ are the two incoming frequencies, the strongest attraction occurs for zero total frequency $\omega_1+ \omega_2$.
 Left plot: Data for the scheme with three fermionic Matsubara frequencies without channel decomposition, Right plot: for the one-frequency parametrization with at-scale channel coupling. All data at $T=0.05t$, $\mu=-0.95t$ and $t'=-0.25t$. }
  \label{CooperFreq}
\end{figure}

Besides the energy scales for the Cooper instability it is also interesting to ask to what precision the frequency dependence of the pairing interaction near the instability can be approximated by the one-frequency scheme.  A comparison between three-frequency flow and the at-scale channel-coupling scheme is shown in Fig. \ref{CooperFreq}. In both cases, only the interaction with zero total incoming frequency diverge. The enhancement of this 'diagonal' feature clearly comes from the particle-particle channel. However, along this diagonal feature, the three-frequency data show a pronounced fall-off to higher values of the fermionic Matsubara frequencies of the incoming lines. This is clearly caused by interplay between external frequencies and summation frequency in the particle-particle loop. Only if the external frequency is small, the large contributions of the loops lines at small summation frequency coincide with small frequency transfer $t$ and thus large values of the phonon-mediated interaction. As the one-frequency schemes only use the initial interaction at a fixed frequency and decouple the external frequencies from the loop frequency, we cannot expect to observe this behavior there. Hence, only the vertex at small external frequencies is approximated well in the one-frequency scheme with at-scale channel coupling. Nevertheless, in this scheme, for the leading interaction components with zero total frequency the decay of the phonon-mediated interaction with the external frequencies is not missed in the particle-particle channel. It is emulated by the 'at-scale' prescription, by using the (usually smaller) value of the phonon-mediated interaction at the $s_0$-frequency equal to the RG scale. This is the reason why the critical scales of the at-scale prescription agree rather well and show the same qualitative behavior as the flows with the full frequency dependence.

\subsection{Phonon-mediated interaction and local repulsion}
\begin{figure}
 \includegraphics[width=.79\columnwidth]{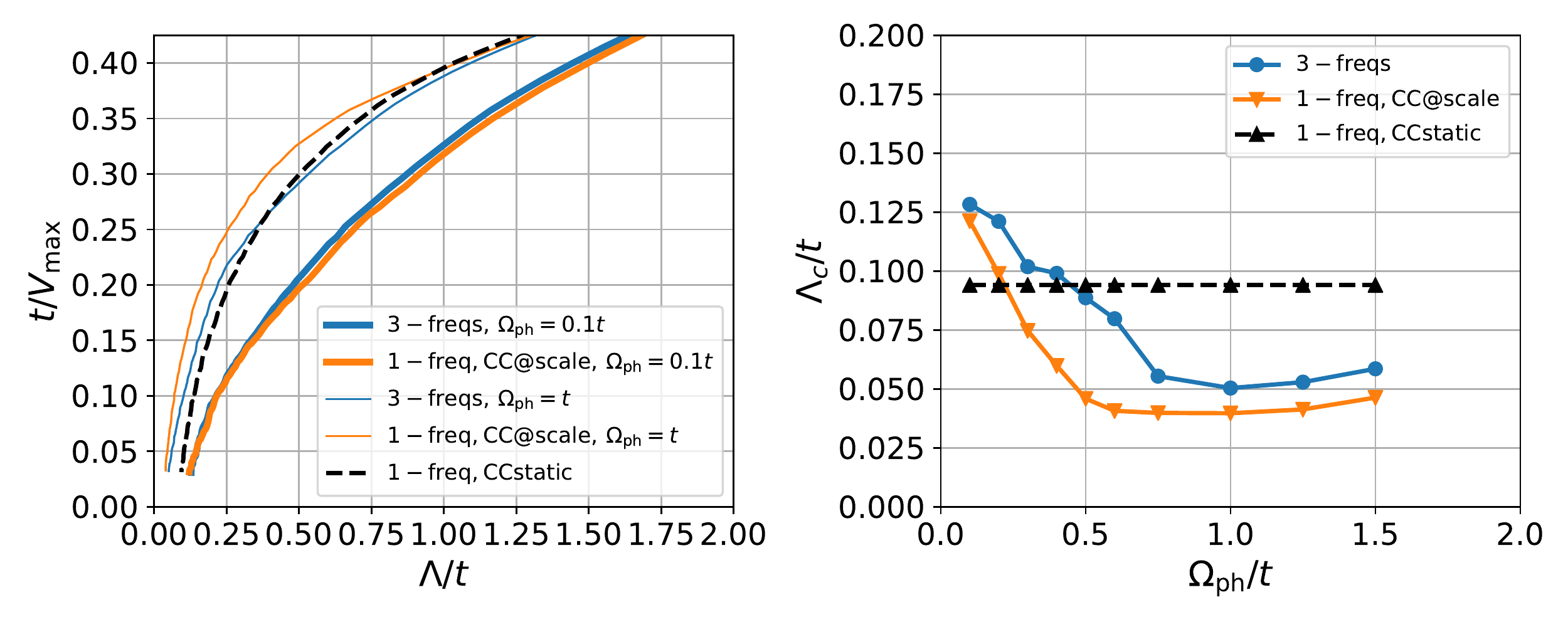}%
 \caption{Flows to strong coupling with all RG channels, with  bare interaction as a sum of a phonon-mediated part with $V_{\mathrm{ph}}=4t$ and a repulsive Hubbard interaction $U=2t$. Left plot: RG flows of the inverse maximal coupling constants for two different phonon frequencies in three different treatments of the frequency dependence, three fermionic Matsubara frequencies without channel decomposition, and the two one-frequency parametrizations with static and at-scale channel coupling. 
 The static channel-coupling results are again independent of the phonon frequency $\Omega$, as only the peak of the phonon-propagator at zero frequency enters in the particle-particle channel.  Right plot: Critical RG scale as function of the phonon frequency for the three schemes. All data at $T=0.05t$, $\mu=-0.95t$ and $t'=-0.25t$. }
  \label{CooperU}
\end{figure}

Next we allow for all one-loop terms on the right-hand side to contribute and add an onsite repulsion $U$ to the initial interaction. Physically, this creates a competition between the local attraction that prefers onsite Cooper pairing and, close to half band filling, charge density wave formation with repulsive tendencies that induce antiferromagnetic tendencies. Results for the critical scales with the three different schemes for the treatment of the frequency dependence are shown in Fig. \ref{CooperU}. In the left plot we show the flows of the inverse maximal coupling strength for two different values of the phonon frequency as function of the flow scale. While the static one-frequency approximation diverges at a rather different scale than the two other schemes, the divergence in the three-frequency scheme and in the at-scale channel-coupling scheme are in better agreement. If we plot the critical scales as function of the phonon frequency, the three-frequency and the at-scale scheme deviate more strongly than in the case without local repulsion, but they follow the same trend.  In contrast, the static channel-coupling approximation does not reflect any change with variation of the phonon frequency. So it is certainly less useful also in this case with partially frequency dependent bare interactions. The at-scale approximation seems to work well on 'semi-quantitative' level, with acceptable errors that lie for many cases in the 20$\%$-range. 

\begin{figure}
 \includegraphics[width=.79\columnwidth]{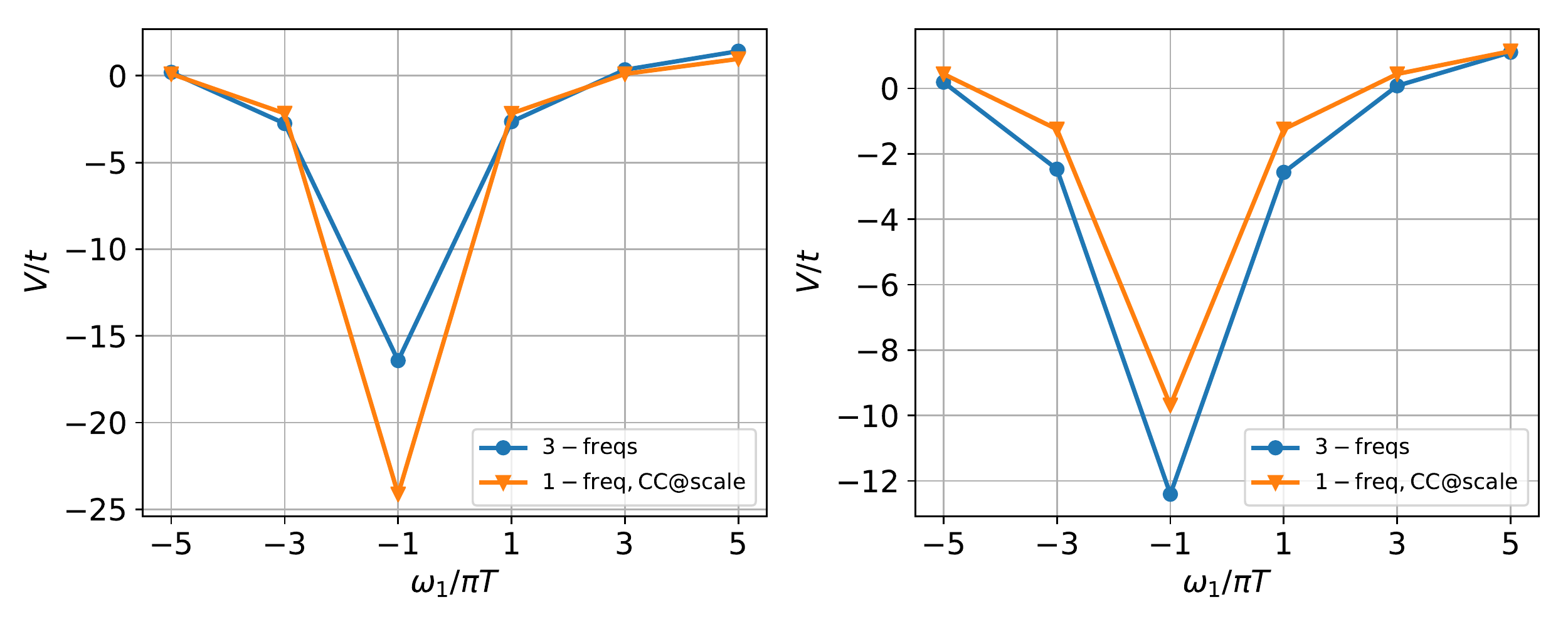}%
 \caption{Frequency dependences of the effective interactions close to the instability with maximal coupling $\sim 30t$ in the three-frequency and the at-scale one-frequency scheme, i.e. at $\Lambda= 0.054t$ for the three-frequency scheme and $\Lambda =0.041t$ for the at scale scheme, with parameters $T=0.05t$, $\mu=-0.95t$ and $t'=-0.25t$, $\Omega_{\mathrm{ph}} =0.6t$. Left plot: Data for $V(35,11,5,\omega_1,\omega_2 = \pi T, \omega_3= -\pi T)$ vs. $\omega_1$. This process has zero total momentum and is enhanced to strong attraction by the Cooper pairing channel.
 Right plot: Data for $V(35,23,2,\omega_1, \omega_2 = \pi T, \omega_3= -\pi T)$ vs. $\omega_1$. This latter situation has momentum transfer $(\pi,\pi)$ in the $D$-channel, i.e. couples to the CDW tendencies. Its total momentum is different from zero, and therefore it is not boosted by the Cooper channel. See Fig. \ref{Fig1points}Êfor the location of the momentum points.
 }
  \label{CooperUFreq}
\end{figure}

In Fig. \ref{CooperUFreq} we also plot cuts through the frequency dependence of the effective interactions obtained in the three-frequency and the at-scale scheme, taken close to the respective critical scales of the same order as the temperature. Again, we can observe clear quantitative differences exceeding 10$\%$. These basically reflect the different critical scales of the schemes. The at-scale scheme has a lower $\Lambda_c$ is more pairing tendencies, leading to a stronger feature in the left plot that has zero total incoming wavevector, while the three-frequency scheme with higher $\Lambda_c$ exhibits still stronger charge-density wave contributions and is more more enhanced at wavevector transfer $(\pi,\pi)$.  
Nevertheless, the gross structures remain the same.  Hence, the at-scale approximation remains a useful tool also for details of the frequency dependence even in this case with quite different (retarded and non-retarded)  competing interactions.

\subsection{Repulsive Hubbard model at van Hove filling}
We can now ask if the at-scale prescription also performs well in the case without any explicit frequency dependence of the bare interaction. In agreement with the standard picture of spin-fluctuation mediated pairing, the magnetic channel (in our case in the form of the $C$-channel) will build up fluctuations around a preferred wavevector, which is $\vec{Q}= (\pi,\pi)$ for our model. This means that $C_\Lambda (\vec{k}_1,\vec{k}_2, \vec{k}_3, u_0) $, starting out from a flat initial condition, will form a peak around $\vec{k}_3- \vec{k}_2 \approx \vec{Q}$ and $u_0 \approx 0$ that gains in magnitude and sharpness for smaller $\Lambda$. Fed back into the particle-particle channel, this will cause $d_{x^2-y^2}$-wave pairing tendencies. If we only feed back the $u_0$-value of $C_\Lambda$, as in the static approximation, we might overestimate the pairing strength. If we use the at-scale prescription, we insert not the peak value of $C_\Lambda$  but a reduced value at $u_0 = \Lambda$. This value should be a good representative for the full Matsubara summation with frequency-dependent interactions included, as the main contributions of the cutoff-carrying loop integrals will come from the frequency values with summation frequency equal to the cutoff and for small external frequencies, the transfer frequencies will lie close to the summation frequency. Again, we can extend this statement from the frequency-cutoff flows to  energy-shell cutoff-flows as the leading loop integrals in both schemes give basically the same. Thus, the peak structure of the spin fluctuation propagator should be reasonably implemented in this at-scale approximation.

\begin{figure}
 \includegraphics[width=.79\columnwidth]{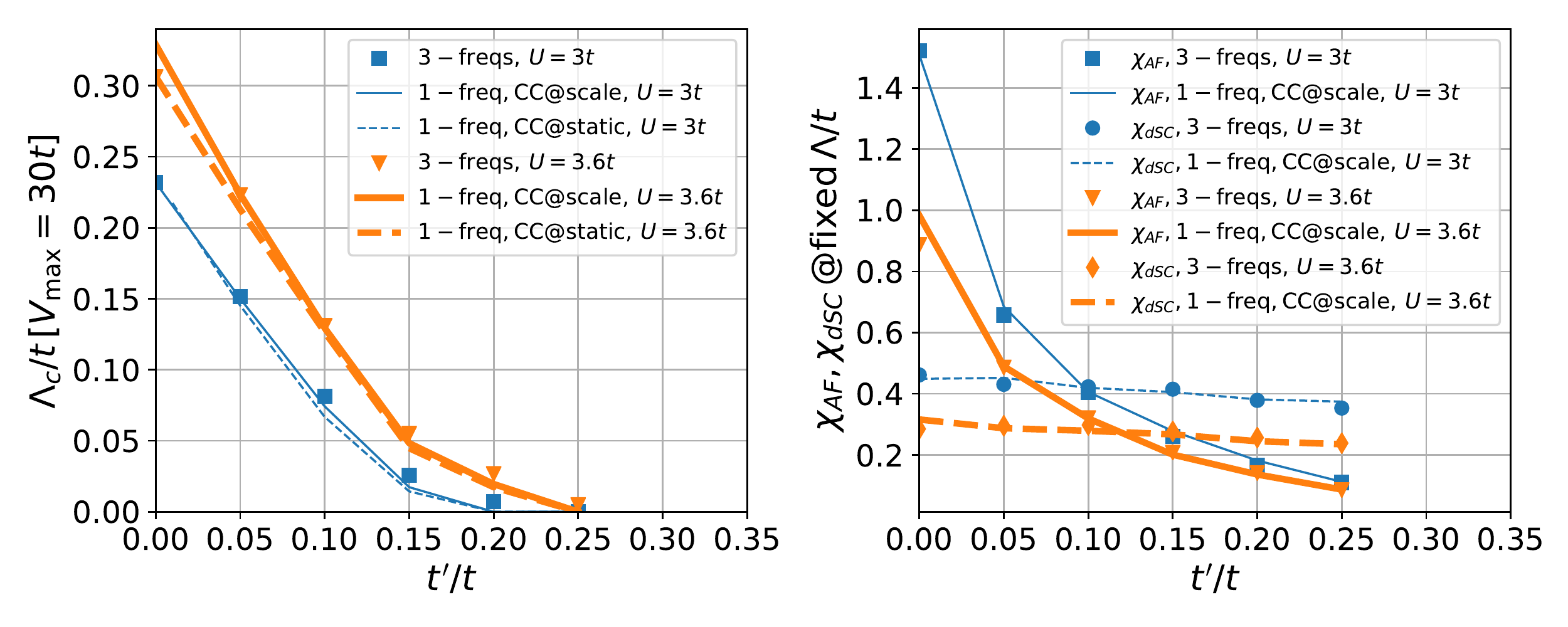}%
 \caption{Left: Critical RG scales for the three-frequency scheme and for the at-scale and static one-frequency parametrizations, for non-retarded Hubbard interactions at the van Hove filling with $\mu=-4t'$ and $U=3t$ or $U=3.6t$, $T=0.05t$, versus second nearest neighbor hopping $t'$. Right: AF spin and $d$-wave pairing susceptibilities at fixed scales $\Lambda=0.25t$ for $U=3t$ and $\Lambda=0.36t$ for $U=3.6t$, again at van Hove filling and versus second nearest neigbor hopping $t'$.}
  \label{Uvh}
\end{figure}

In Fig. \ref{Uvh} we show data for the fRG with the three-frequency parametrization and the at-scale and static channel-coupling one-frequency approximations for various values of the next-nearest neighbor hopping $t'$ at the van Hove filling. This trajectory in the parameter space has been used in numerous other works\cite{tflow,husemann,giering,lichtenstein}. For small $t'$ the leading instability is in the AF channel. Upon increasing the $t'$ and hence the Fermi surface curvature, $d$-wave pairing takes over as leading instability. This interplay has been analyzed in the previous works and here we do not discuss it further. We also note that while our data here agree qualitatively with previous works, we do not expect quantitative agreement, as the energy-shell cutoff used here is blind with respect to ferromagnetic fluctuations and as we only use a coarser momentum-space patching with only 16 points around the Fermi surface.

The upshot of the data in Fig. \ref{Uvh} is that the at-scale approximation performs quite well compared to the more precise three-frequency parametrization. Both, the critical scales for two different values of $U$ as well as the two most relevant susceptibilities in the right plot agree up to a few percent in most cases and give the same physical picture. The highest relative disagreement occurs, quite understandably,  when the critical scales are of the same order or smaller than the temperature ($T=0.05t$ in this case), i.e. when the end point of the flow puts the system in the critical region.

\begin{figure}
 \includegraphics[width=.79\columnwidth]{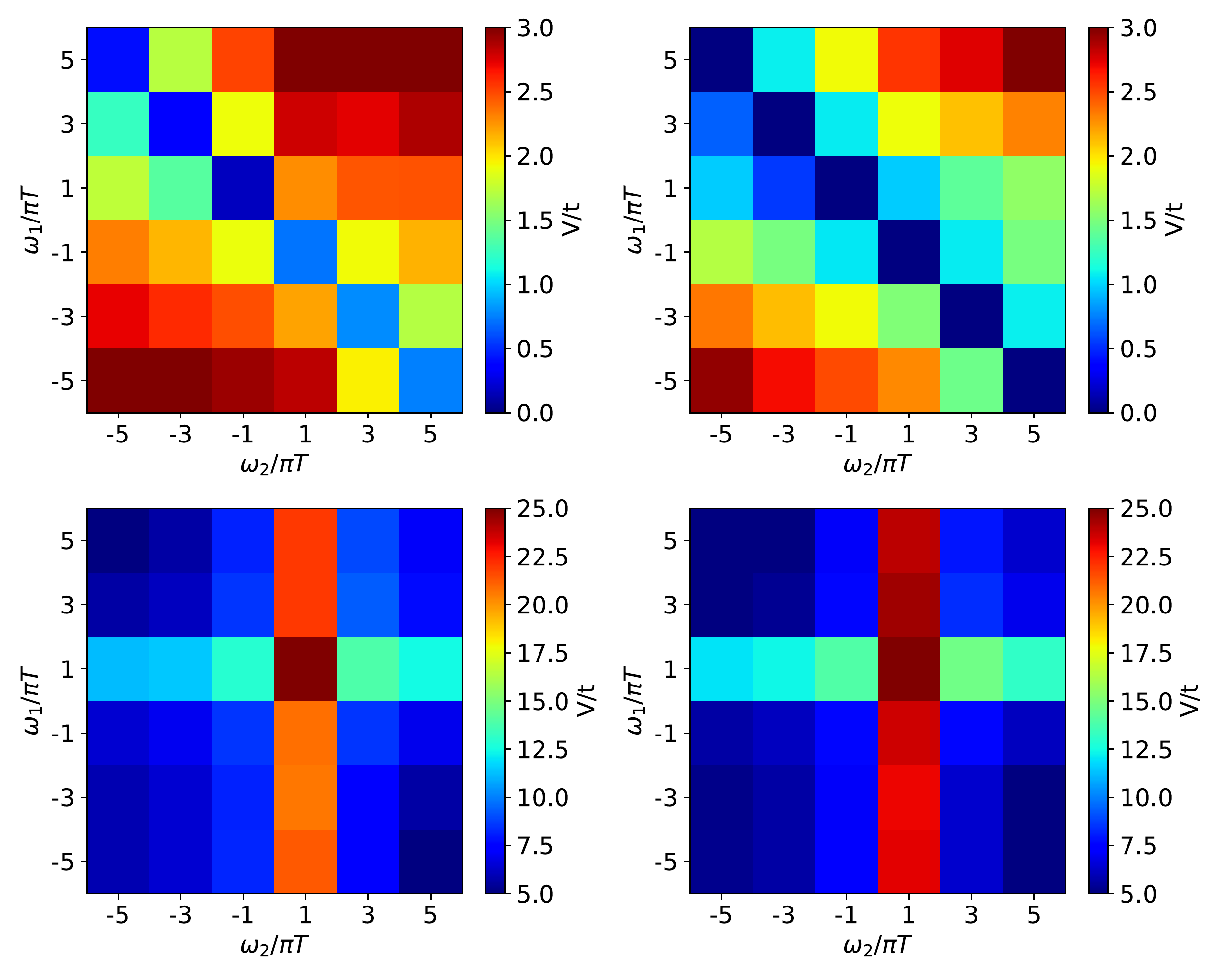}%
 \caption{Frequency dependence of the diverging interaction for the Hubbard model for $t'=-0.15t$ and van Hove filling, $U=3t$, $T=0.05t$, with the first outgoing leg at wavevector point $2$ and first outgoing frequency $\omega_3 = \pi T$. The colorbars denote the strength of the interaction. $\omega_1$ and $\omega_2$ are the two incoming frequencies.
 Left plots: Data for the scheme with three fermionic Matsubara frequencies without channel decomposition. Right plots: Data for the one-frequency parametrization with at-scale channel coupling.  For the upper plots, the incoming wavectors are at points 47 and 23, i.e. have zero total incoming momentum and the Cooper pair scatters within one saddle point. For the lower plots, the incoming points are both at 31 and the first outgoing is at 5, which means the particles are scattered in an Umklapp process by roughly $(\pi,\pi)$ across the Fermi surface.}
  \label{UvhFreq}
\end{figure}

In Fig. \ref{UvhFreq} we compare the frequency dependences obtained at the respective critical scale with the three-frequency parametrization on the left with that from the at-scale one-frequency approximation on the right. In upper and lower panels, the qualitative agreement is well visible. In the upper plots we show  a wavevector combination with zero total incoming wavevector, i.e. processes that contribute to the Cooper pairing channel for zero total frequency. 
The lower plots are for Umklapp processes across the at $t'=-0.15t$  imperfectly nested Fermi surface. On the quantitative level, some differences can be seen. For instance, the three-frequency data in the upper left plot is somewhat more repulsive than the one-frequency data in the upper right one, while its is opposite for the lower plots.  In the upper panels, the Cooper pair is scattered only by a small wavevector. This process belongs to the attractive part of the $d$-wave scattering and the effective interaction is smaller than the bare interaction in this data, in particular for the diagonal line that belongs to zero total frequency. For these values to become really negative, we would have to got to lower $T$. Compared to this data in both upper plots for small Cooper pair scattering wavevector, the values in the lower panels, corresponding to Umklapp processes with wavevector transfer near $(\pi,\pi)$, are much more repulsive. Furthermore, no diagonal suppression feature for zero total momentum is seen. Instead, the lines with $\omega_2-\omega_3=0$ and $\omega_1-\omega_3=0$ are strongly enhanced. This is because both wavevector transfers $\vec{k}_{1/2}-\vec{k}_3 \approx (\pi,\pi)$ and these scattering processes are enhanced by the AF fluctuations. These are dominantly present in the $C$-channel and in the $D$-channel with roughly half of the strength.  

Hence, also in this case the at-scale approximation provides a useful simplification. Actually for the cases shown here, the static approximation would also remain close to the two other schemes. However, in the cases with stronger frequency dependence as shown in the previous subsection, the static approximation performed certainly less  convincing. Hence the clear recommendation goes to the at-scale approximation, as it is potentially also sensitive to developing interactions with stronger frequency dependences.

\section{Discussion and conclusions}
We have discussed two possible simplifications of the frequency dependence of the two-particle interaction vertex in the functional renormalization group formalism for fermions on two-dimensional lattices. The benchmark approximation is the brute-force treatment of the full frequency dependence in terms of three independent fermionic Matsubara frequencies for the two incoming and one outgoing leg of the vertex.  This produces a vertex size that grows like the 3rd power of the number of frequencies $N_\omega$ used and a numerical effort that grows even more strongly (quartic in the number of frequencies in naive estimation). We ran this scheme for $N_\omega =6, 8$ or $10$.
We have contrasted the results of these flows  with two numerically much lighter one-frequency parametrizations\cite{karrasch} (with the first also used in Refs. \onlinecite{bauer,weidinger,markhof}) of the vertex frequency dependence. They make use of the channel decomposition that was tested quite successfully for the momentum dependence\cite{husemann,giering,wang, lichtenstein,sanchez}.  Besides increasing the vertex size only linearly in the number of frequencies, these approximation have the additional benefit that the frequency dependence of the vertex is pulled out of the loop frequency summations, leading to an additional numerical speedup in standard cases. While the so-called static channel-coupling approximation was seen to fail qualitatively in the case of retarded interactions, the one-frequency scheme with at-scale approximation performed well also when a phonon-mediated interaction with strong retardation was present. We showed the approximate agreement of the critical scales in the at-scale scheme compared to the three-frequency scheme. Similar agreement was found when considering the detailed frequency dependences of the vertex and in the relevant susceptibilities. As the remaining frequency dependence of the interactions in the various channels follows that on the corresponding one-loop bubbles, strange instabilities of the interactions at finite frequencies as found in Ref. \onlinecite{husemann} and discussed to be spurious in Ref. \onlinecite{vilardi}, or other problematic frequency dependences should not occur in this scheme. Hence, for a qualitative and semi-quantitative (i.e. if one is not interested in a precision of a few percent), the at-scale one-frequency approximation appears to be a viable tool for future studies. 

In the current study the  one-frequency parametrization with static channel coupling does not perform well in the comparisons with explicit frequency dependence of the initial interaction. If this dependence is absent however, e.g. for Hubbard initial interactions only, the critical scales of the static channel-coupling scheme were found to be close to those of the three-frequency parametrization, often within 10$\%$ if the critical scales are still higher than the temperature. The critical scales with static channel coupling can be argued to be identical to those without any frequency dependence of the interactions. The latter approximation was used in numerous previous fRG studies of two-dimensional systems\cite{metzner,platt}. Our study shows that at least for frequency-independent interactions the critical scales found in those works should serve as reasonable estimates. 

We close by remarking that if one wants to use a one-frequency parametrization of the interactions, the at-scale approximation appears to the safest option with the same numerical cost as the static channel coupling. If one wants to improve on that, testing form factor decompositions also on the frequency axis seem worthwhile\cite{yirga}. The benefits of all these approximations are a much reduced numerical effort compared to the full schemes with non-decomposed three-frequency dependence. This opens many other possibilities like larger frequency window, lower temperature, higher momentum resolution or more bands to be considered. Notably, these approximations do not spoil the matrix-multiplication structure of the right hand side of the flow equations of the vertex that allowed for a well-scaling parallelization within the truncated-unity fRG formalism\cite{lichtenstein,sanchez}. 
Having at hand an access to frequency-resolved interactions, the computation of meaningful and physically important selfenergies should be feasible without much additional effort. This issue will be explored next. Furthermore, the door seems open for studies of the interplay of phonons and electronic interaction beyond the Migdal-Eliashberg level, i.e. taking into account vertex corrections. \\[1mm]

We acknowledge discussions with N. Dittmann, C. Eckhardt, J. Ehrlich, K. Eissing, C. Hille, D. Kennes, L. Markhof,  A. Tagliavini, D. Rohe, G. Schober, D. Vilardi and  N. Yirga. We received support through the DFG research training group 1995.

\bibliography{freqpap}

\end{document}